\documentclass[aps,amsmath,amssymb,twocolumn]{revtex4-1} \usepackage{epsfig}
\usepackage{color} 

\begin{document}

\title{Function changing mutations in glucocorticoid receptor
  evolution correlate with their relevance to mode coupling}
\author{Batuhan Kav$^{1,2}$, Murat \" Ozt\" urk$^1$ and Alkan Kabak\c
  c\i o\u glu$^1$} \affiliation{$^1$Department of Physics, Ko\c c
  University, Sar\i yer, 34450, Istanbul, Turkey\\ $^2$Max Planck
  Institute for Colloids and Interfaces, Science Park Golm, 14424,
  Potsdam, Germany}
%% \author{Alkan Kabak\c c\i o\u glu\thanks{akabakcioglu@ku.edu.tr}} \affiliation{Department of Physics,
%% Ko\c c University, Sar\i yer, 34450, Istanbul, Turkey}

\begin{abstract} Nonlinear effects in protein dynamics are expected to play role
in function, particularly of allosteric nature, by facilitating energy
transfer between vibrational modes. A recently proposed method
focusing on the non-Gaussian shape of the population near equilibrium
projects this information onto real space in order to identify the
aminoacids relevant to function. We here apply this method to three
ancestral proteins in glucocorticoid receptor (GR) family and show
that the mutations that restrict functional activity during GR
evolution correlate significantly with locations that are highlighted
by the nonlinear contribution to the near-native configurational
distribution. Our findings demonstrate that nonlinear effects are not
only indispensible for understanding functionality in proteins, but
they can also be harnessed into a predictive tool for functional site
determination.  \end{abstract}

\maketitle

\newpage \subsection{Introduction} \label{sec:intro} Mechanisms of
information transfer and function in proteins continue to be
challenging problems where different points of view compete. The
ensemble view, i.e., that a ligand binding event triggers allostery by
modifying the free energy landscape is now a commonly recognized
paradigm~\cite{kumar2000folding,tsai1999folding,okazaki2008dynamic}.
The so called ``population shift'' picture is helpful in understanding
allostery without shape change and finds support from recent NMR
studies~\cite{kar2010allostery}.
%% Change in spatial configurations of % signaling proteins NtrC, CheY, Rap2a,
%Cdc42, and IRK; adenylate kinase % enzyme and hemoglobin upon ligand binding
%are just a few of many % examples~\cite{volkman2001two, kar2010allostery,
%arora2007large, %   benesch1967effect}.  
Models which focus on mechanistic aspects, such as the suppression of
a certain vibrational mode or energy transfer between two
modes~\cite{brooks1983harmonic,haliloglu1997gaussian,yogurtcu2009statistical,ma2005usefulness,piazza2009long}
are also widely employed, thanks to the intuitive picture they
offer. Through such models, it is, for example, possible to discuss
positive/negative allostery~\cite{rodgers2013modulation}.
%% and identify residues that alter the dynamics.

A recently proposed approach that rests on dynamical simulations
around equilibrium delivers a tool for predicting locations relevant
to mode
coupling~\cite{kabakcciouglu2010anharmonicity,varol2014mode}. The
central idea of the method is to quantify the nonlinear contribution
(required for energy transfer between vibrational modes) to the
near-native configurational probability distribution function and
identify the residues on which it has the highest impact. Present work
is an application of this method to a set of ancestral glucocorticoid
receptor (GR) proteins and observes a statistically significant
correlation between locations underlined by mode coupling and function
changing mutations that took place during the evolution of the GR
proteins. Multiple molecular dynamics (MD) trajectories of $\approx
0.1 \mu s$ for each protein further allow us to discuss the robustness
of the method between independent MD runs, as well as the sensitivity
of our findings to the simulation length.

\subsection{Extracting information on mode coupling} \label{sec:method} Consider
a protein composed of $N$ aminoacids. Let the Cartesian coordinates of
carbon-alpha ($C_\alpha$) atoms be stored in the vector $\boldsymbol R$ of
length $3N$, encoding a coarse representation of the protein's spatial
arrangement, or configuration. The configurational probability distribution,
$p(\boldsymbol{R})$ can be derived numerically from an ${\cal M} \times 3N$
real-valued matrix, where ${\cal M}$ is the number of snapshots acquired by
sampling sufficiently long MD trajectories. This configurational distribution is
then used to determine fluctuations around the mean structure,
$\boldsymbol{\delta R} = \boldsymbol{R}-\langle\boldsymbol{R}\rangle$, where
$\langle\cdot \rangle$ indicates averaging over time and multiple MD
trajectories generated using different random seeds.

Within the framework of elastic network models, the configurational distribution
is most conveniently expressed in terms of ``modal fluctuations''
\begin{equation} \label{eq:gamma} \boldsymbol{\delta
r}=\Gamma^{-1/2}\boldsymbol{\delta R} \end{equation} where
$\Gamma=\langle\boldsymbol{\delta R}\boldsymbol{\delta R}^{T}\rangle$ is the
covariance matrix associated with real-space fluctuations. A general analytical
expression for $p(\boldsymbol{\delta r})$ in terms of Hermite tensor polynomials
was originally proposed by Flory~\cite{flory1974moments}: \begin{equation}
\label{eq:flory} p(\boldsymbol{\delta r})=\frac{e^{-\delta
r^2/2}}{(2\pi)^{3N/2}}\sum_{\nu=0}^{\infty} C_\nu
\boldsymbol{H}_\nu(\boldsymbol{\delta r})\ , \end{equation} where
$\boldsymbol{H}_\nu$ denotes the Hermite tensor polynomial of rank $\nu$, and
the coefficients $C_\nu$ follow from the orthogonality relation $\int
\boldsymbol{H}_\nu(\boldsymbol{\delta r})\boldsymbol{H}_\mu(\boldsymbol{\delta
r}) d\boldsymbol{r} = (\nu!)^{3N}\,\delta_{\nu,\mu}$.

In $\{\boldsymbol{r}\}$ basis, all expansion coefficients $C_{\nu\neq 0}$ in
Eq.(\ref{eq:flory}) vanish for a perfectly elastic network. As a result, the
configurational distribution of such a linear system is separable into $3N$
identical Gaussian functions with unit standard deviation and zero mean:
\begin{equation} \label{gaussian} p^{(g)}(\boldsymbol{\delta r}) =
\prod_{m=1}^{3N} \frac{\exp[-(\delta r_{m})^{2}/2]}{\sqrt{2\pi}}\ .
\end{equation} The superscript $^{(g)}$ implies the Gaussian product form of
$p$. All vibrational modes of the linear system are represented in an identical
fashion in this normalized form of the distribution. Yet, their physical
difference is evident from and encoded in the corresponding eigenvalues and
eigenvectors of $\Gamma$. 

Nonlinearity can be introduced into this picture {\em without coupling the
vibrational modes}, by simply adding higher-order terms to the corresponding
harmonic oscillator Hamiltonians: \begin{equation} \label{eq:hamiltonian} {\cal
H}_m = \frac{p_m^2}{2\mu_m} + \frac{1}{2}k_m(\delta r_m)^2 +
\sum_{n>2}\alpha_{n,m} (\delta r_m)^n \end{equation} where $\mu_m$ and $k_m$ are
the effective mass and the spring constant for the $m^{th}$ mode and
$\alpha_{n,m}$ indicate the strength of higher-order terms in appropriate units,
all of which can in principle be derived from the underlying dynamics. The
resulting configurational distribution function can then be expressed as
\begin{equation} \label{eq:marginal_anh} p^{(s)}(\boldsymbol{\delta r}) =
\prod_m p_m(\delta r_m)\ .  \end{equation} where \begin{equation} \label{eq:pm}
p_m(\delta r_m) = \frac{\exp[-(\delta r_{m})^{2}/2]}{\sqrt{2\pi}} \bigg[1 +
\sum_{\nu=3}^{\infty}c^{m}_{\nu}H_{\nu}({\delta r}_{m}) \bigg]\ .
\end{equation}

Note that, $H_\nu$ above is now the ordinary Hermite polynomial of rank $\nu$.
Eqs.(\ref{eq:marginal_anh},\ref{eq:pm}) describe the most general separable
distribution for the variables $\{\delta r_m\}$, hence the superscript $^{(s)}$.

An arbitrary configurational distribution can be expressed in a similar fashion,
starting from Eq.(\ref{eq:flory})~\cite{kabakcciouglu2010anharmonicity}:
\begin{widetext} \begin{equation} \label{eq:hermite} p(\boldsymbol{\delta
r})=\frac{1}{\sqrt{(2\pi)^{3N}}}e^{-\sum_{m}{\delta r}_{m} ^{2}/2} \
\bigg[1+\sum_{m}\sum_{\nu=3}^{\infty}c^{m}_{\nu}H_{\nu}({\delta r}_{m})+ \\
\sum_{m \ne
n}\sum_{\nu=3}^{\infty}\sum_{\mu=1}^{\nu-1}c_{\mu,\nu-\mu}^{m,n}H_{\mu}({\delta
r}_{m}) H_{\nu-\mu}({\delta r}_{n})+\sum_{m \ne n \ne p} \dots \bigg]
\end{equation} \end{widetext} Eq.(\ref{eq:hermite}) allows one to address all
nonlinear corrections to the Gaussian description of near-native fluctuations in
a methodical, order-by-order fashion. For the type of nonlinearity which yields
Eq.(\ref{eq:marginal_anh}) in a suitably chosen basis $\{\boldsymbol{r}\}$ we
propose the term ``marginal anharmonicity'', because it is fully characterized
by marginal distributions $p_m(\delta r_m)$. In other words, all coefficients $
c^{m,n}_{\mu,\nu-\mu}$ and their higher-order counterparts in
Eq.(\ref{eq:hermite}) can be derived in this case from the coefficients of
``one-body'' terms $c^{m}_{\nu}=\langle H_{\nu}({\delta r}_{m})\rangle/\nu!$.
For example, $c_{\mu,\nu-\mu}^{m,n}=\frac{\nu!}{(\nu-\mu)!}\,c^{m}_\mu
c^{n}_{\nu-\mu}$. On the other hand, the diagonalization problem, that is,
determining the transformation $\boldsymbol{R} \to \boldsymbol{r}$ for an
arbitrary marginally anharmonic system is significantly harder than that for a
linear one. (This problem arises naturally in signal processing field and is
known as ``Independent Component Analysis'' or
ICA~\cite{hyvarinen2004independent}.)

Deviations in $p(\boldsymbol{r})$ from the separable form in
Eq.(\ref{eq:marginal_anh}) are, by construction, due to coupling between
(possibly anharmonic) vibrational modes. In fact, when such coupling is strong,
the solutions of the nonlinear system will depart significantly from those found
in the noninteracting limit in Eq.(\ref{eq:marginal_anh}); so that, describing
the dynamics around the vibrational ``modes'' in Eq.(\ref{eq:hamiltonian}) may
no more be justified. However, numerous earlier studies suggest that a
quasi-elastic treatment of protein dynamics is an adequate and fruitful course
at physiological
temperatures~\cite{brooks1983harmonic,kurplus1983dynamics,levy1982molecular,levitt1985protein,doruker2000dynamics}.
Therefore, in the following we assume that the eigenvectors of $\Gamma$ in
Eq.(\ref{eq:gamma}) continue to serve as a meaningful basis in which $p_m(r_m)$
calculated along the eigen-directions yield a good approximation for the best
marginally anharmonic description $p^{(s)}$ of the system.  Our results reported
below corroborate this expectation.

While the degree of coupling between mode pairs (also triplets, quartets and so
on) can be discussed by investigating corresponding high-order terms in
Eq.(\ref{eq:hermite}), it was shown earlier that a cumulative treatment of all
coupling corrections to Eq.(\ref{eq:marginal_anh}) is not only possible but also
easier~\cite{varol2014mode}. This is done by quantifying the difference between
the near-native conformational distribution obtained from MD (which, in
principle, includes effects at all orders) and the best description of the data
in terms independent modes, as given by Eq.(\ref{eq:marginal_anh}).

A motivation for studying mode coupling is to gain insight about the mechanism
and regulation of function in allosteric proteins. Accordingly, one is typically
interested in locations on the structure where mutations or immobilization by
means of ligand binding alter protein's activity. After isolating multi-mode
contributions to fluctuations as described above, the next step is therefore to
project this information onto the protein structure in order to identify regions
which are critical for mediating the coupling between relevant modes. To this
end, we first map the distribution $p^{(s)}$ back onto Cartesian space by
\begin{equation} p^{(s)}(\boldsymbol{\delta R}) = p^{(s)}(\boldsymbol{\delta
r})/\sqrt{\det\,\Gamma}\ .  \end{equation} Next, the difference between
$p(\boldsymbol{\delta R}_i)$ and $p^{(s)}(\boldsymbol{\delta R}_i)$ is measured
along each Cartesian component of the position vector $\boldsymbol{R}_i$ of
residue $i$ by means of the Jensen-Shannon (JS) divergence $d_{js}[p,p^{(s)}]$
defined as~\cite{manning1999foundations}: \begin{equation} \label{js}
d_{js}[p,p^{s}] = \frac{1}{2}\,\big[ d_{kl}(p,M) +d_{kl}(p^{s},M) \big]
\end{equation} where $M=\frac{1}{2}(p+p^{s})$ and $d_{kl}(p,q)$ is the
Kullback-Leibler divergence that is given by \begin{equation}
d_{kl}(p,q)=\int_{-\infty}^{\infty}p(x) \, \ln\bigg[\frac{p(x)}{q(x)}\bigg] dx \
.  \end{equation} An advantage of using JS divergence is that it is symmetric,
therefore immune to the possibility that one of the two distributions may vanish
at certain points (Kullback-Leibler divergence yields infinity for such
instances). 

Following the recipe above, we calculate the mode-coupling score for each amino
acid in a protein, as the sum of JS divergences calculated along the components
of $\boldsymbol{R}_i$. Note that, comparing $p$ and $p^{(s)}$ aminoacid by
aminoacid rather than in their full domain not only yields a testable output
(locations relevant to mode coupling in the protein), but is also more robust to
stochastic fluctuations inherent to the method, simply because of reduced
dimensionality. That is, histograms for spatial fluctuations of single
aminoacids can be represented with much fewer bins compared to joint
distributions $p$ or $p^{(s)}$ for the whole protein, therefore they are
sufficiently well sampled along the MD trajectories.  Variations in the output
as a function of the simulation length as well as between independent MD runs
are investigated in Section~\ref{sec:md}.

An earlier version of this prescription was applied to motor protein myosin II
with encouraging results~\cite{varol2014mode}. We here consider the ancestral
chain of GR proteins and show that the outlined analysis reveals a significant
bias in mode-coupling scores for experimentally validated function-altering
mutations in the family.

\subsection{Glucocortocoid receptors and evolutionary data} \label{sec:gr}
Glucocorticoid receptors are a class of endogenous steroid hormones that
regulate imflammatory and stress responses, growth, development, and
apoptosis~\cite{kino2009glucocorticoid, nicolaides2010human,
chrousos2004glucocorticoid, clark, zhou2005human, barnes1998anti,
sapolsky2000glucocorticoids}. GR positively regulates transcription through a
process known as transactivation in which the ligand-bound GR dislocates from
cytoplasm to enter cell nucleus where it activates
transcription~\cite{carson2014glucocorticoid}. Its paralogous counterpart,
mineralocorticoid receptor (MR) is mainly responsible for regulating electrolyte
homeostasis~\cite{li2005structural}. While GR binds glucocorticoid hormone
cortisol\cite{bentley1998comparative}, MR acts as a host for aldosterone,
11-deoxycorticosterone (DOC) and with a lesser affinity for
cortisol~\cite{yang2009mineralocorticoid}.  Through phylogenetic analysis, the
sequence and the crystal structure of their common ancestor AncCR, as well as
the ancestral GR proteins in cartilaginous fish (AncGR1) and in bony vertebrates
(AncGR2) have been determined~\cite{bridgham2009epistatic, ortlund2007crystal}.
It has been shown that AncCR, similar to AncGR1, indiscriminately binds to DOC,
aldosterone and cortisol. On the other hand, AncGR2 exclusively binds cortisol
and is not activated by aldosterone and DOC.

Considerable effort has been devoted to understanding the basis of ligand
specificity in the evolution of GR~\cite{ortlund2007crystal,
bridgham2009epistatic,harms2014historical}. Structural variations are minute,
with $<1\AA$ RMSD difference between AncGR1 and
AncGR2~\cite{glembo2012collective}. Among the 36 mutations that transform AncGR1
to AncGR2, it has been shown that two strictly conserved mutations, S106P and
L111Q (group X), are sufficient to increase cortisol specifity.  S106P changes
the architecture of ligand binding pocket and allows L111 to be located at a
closer position to the ligand. The effect of L111Q is biochemical rather than
mechanical, since it creates an additional hydrogen bond between 111Q and the
cortisol, which lacks in DOC and aldosterone binding.  Three additional
mutations, L29M, F98I and S212$\delta$ (group Y), wipe out the affinity towards
DOC and aldosterone. However, AncGR1+X+Y structure cannot activate transcription
due to the damaged hydrogen bond network which destabilizes the
activation-function helix.  Two further mutations, N26T and Q105L (group Z), are
necessary in order to reestablish the hydrogen bond network and thereby
stabilize the structure. All together, AncGR1+X+Y+Z, captures AncGR2 phenotype.
A recent study found that XYZ mutations correlate with a measure based on the
difference in fluctuation amplitudes of a residue in principal vibrational modes
of ancestral GR proteins~\cite{glembo2012collective}.

Besides historically occuring mutations, another study on alternative
evolutionary pathways that restore AncGR2 phenotype~\cite{harms2014historical}
demonstrated that, among a set of suggested alternatives, only the mutation pair
Q114L/M197I recovers cortisol sensitivity similar to the historical set of
permissive mutations, albeit with a loss of associated transcriptional function.

\subsection{Specifics on molecular dynamics simulations} \label{sec:md} Crystal
structures of AncCR, AncGR1, and AncGR2 are publicly available at PDB with
accession codes 2Q3Y, 3RY9, and 3GN8, respectively. MD simulations of each were
carried out on Tesla K20 GPUs by means of Amber 14 Molecular Dynamics
Package~\cite{case2014amber} using $ff14sb$
force-field~\cite{cornell1995second}. Ligand molecules were also included in the
simulations and their parametrization were done with Antechamber using
generalized Amber force field~\cite{wang2004development, wang2006automatic}. All
simulations were performed in $(N,P,T)$ ensemble with explicit water solvent and
with Langevin dynamics which maintained the temperature at 310 K and the
pressure at 1 bar. No rigid bonds were assumed. 1 fs timesteps were used between
successive frames while trajectories were captured every 1000 frames, i.e. 1 ps
apart, throughout 128 ns long simulations performed for each sample. Each
protein was simulated four times with the same initial condition but different
random number generator seeds. Before further analysis, trajectories were
aligned by means of the backbone $C_\alpha$ atoms. Discarding the first 7 ns of
each simulation for equilibration, this protocol resulted in $5\times 10^5$
snapshots, derived from $\approx 500\, ns$ long simulations of the near-native
dynamics for each protein.

\subsection{Mode coupling in GR proteins and comparison with evolutionary data}
\label{sec:results} 
\begin{figure}[t] \hspace*{-1cm} \centering
\includegraphics[scale=0.27]{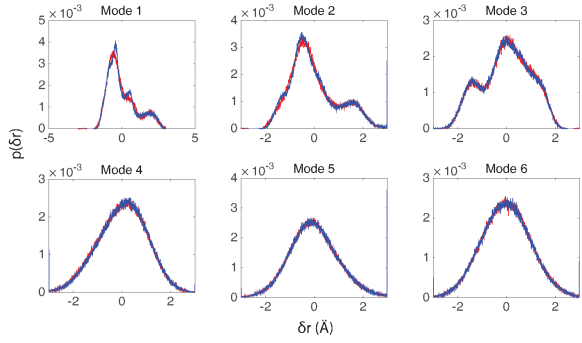} \caption{Marginal distributions for
the amplitudes of six slowest modes of AncGR2. Anharmonicty is most discernible
in the first three and gradually disappears for faster modes.  Analytical
approximations derived from Eq.(\ref{eq:pm}) with a cut-off at $v =32$ are also
shown in red.} \label{fig:marginal} \end{figure}

A substantial contribution from marginal anharmonicity is
observable in the amplitude distributions of the slowest modes of AncGR2 protein
shown in Fig.~\ref{fig:marginal}. Strong deviation from Gaussian in
 eigenmodes with large
eigenvalues is typical for the whole GR family, in fact for most
proteins~\cite{roh2005onsets,hayward1995harmonicity}. A comparison of the
marginal distributions $p_m(\delta r_m)$ and Eq.(\ref{eq:pm}) with a cut-off at
$v=32$ confirms that they are represented accurately by the analytical approach
in Section~\ref{sec:method}.

\begin{figure}[b] \centering \includegraphics[scale=0.24]{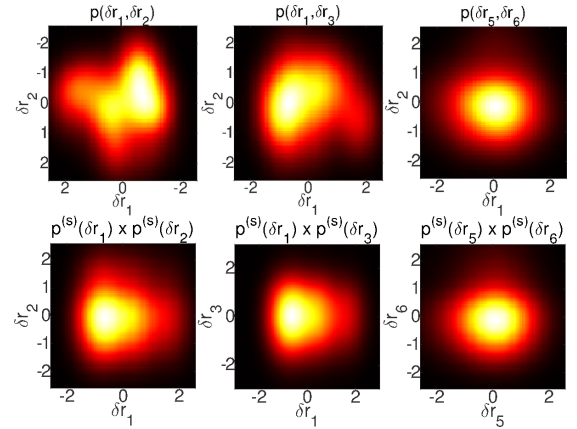}
\caption{Pairwise joint probability distributions given as heat maps for the
amplitudes of three slowest modes in AncGR2 (high probability regions are shown
in yellow). First row corresponds to the MD data for pairs 1-2, 1-3, and 5-6,
respectively. Second row gives the product of corresponding marginal
distributions.  It is evident that, joint distributions of slow modes can not be
captured by Eq.(\ref{eq:marginal_anh}), meaning mode-coupling corrections must
be included.} \label{fig:joint} \end{figure}

On the other hand, presence of mode coupling is evident from the difference
between $p$ and $p^{(s)}$, as shown in Fig.~\ref{fig:joint} by considering joint
amplitude distributions for $(\delta r_m, \delta r_n)$ corresponding to slowest
mode pairs $(m,n)=(1,2),(1,3),(2,3)$.  Contribution of mode coupling is
exemplified by the difference between the two rows of Fig.~\ref{fig:joint}.
Information content of such deviation from marginal anharmonicity in protein
dynamics and its relevance to protein's biological function is our focus in this
study.

\begin{figure*} \includegraphics[width=\textwidth]{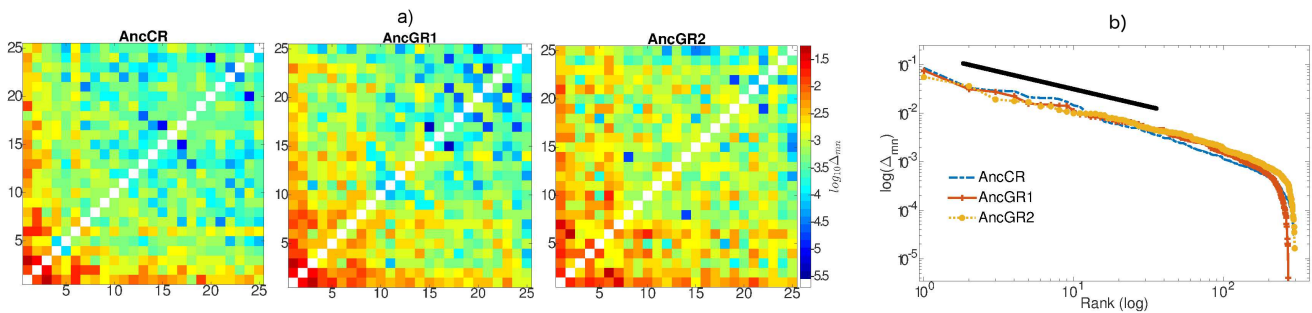}
\caption{(a) Pairwise mode coupling scores for slowest 25 modes of the three
members of GR protein lineage. (b) Mode-coupling scores $\Delta_{mn}$ of
rank-ordered mode pairs for the three proteins. The straight line segment
(black) corresponds to a power-law decay with an exponent $-0.8$. }
\label{fig:pairs} \end{figure*}

We start by asking whether certain mode pairs stand out in the above analysis.
One can assess the overall impact of a mode pair by considering the
(dimensionless) conformational free energy \begin{equation} F[p] = -\int
p(\boldsymbol{\delta r}) \ln p(\boldsymbol{\delta r}) d\boldsymbol{\delta r} =
\frac{1}{\cal M}\,\sum_{i=1}^{\cal M} \ln p(\boldsymbol{\delta r}_i)
\end{equation} calculated with and without the mode coupling contribution from
the mode pair $(m,n)$ in Eq.(\ref{eq:hermite}). For this purpose, we
approximated $p(\boldsymbol{\delta r}_i)$ by the one- and two-body terms spelled
out in Eq.(\ref{eq:hermite}). We then defined $p_{-mn}(\boldsymbol{\delta r}_i)
= p(\boldsymbol{\delta r})\big|_{c^{mn}_{\mu,\nu-\mu}=c^m_\nu c^n_{\nu-\mu}}$,
in which the mode-coupling contribution of the pair $(m,n)$ is discarded. The
difference $\Delta_{mn} = F[p_{-mn}] - F[p]$ is a measure of the impact of the
interaction between modes $m$ and $n$ on protein's behavior near equilibrium.
$\Delta_{mn}$ for all pairs composed out of slowest 25 modes of each protein are
given in Fig.~\ref{fig:pairs}. It is interesting that $\Delta_{mn}$ displays a
power-law dependence with a scaling exponent $\sim -0.8$ over more than two
decades on the rank order of the pair $(m,n)$ (Fig.~\ref{fig:pairs}b). An
exhaustive analysis over all mode pairs was not performed due to its heavy
computational cost.

%After approximating $p(\boldsymbol{\delta r}_i)$ by the one- and two-body terms
%spelled out in Eq.(\ref{eq:hermite}), we define $p_{-mn}(\boldsymbol{\delta r})
%= p(\boldsymbol{\delta r})\big|_{c^m_\nu=c^n_\nu=c^{mn}_{\mu,\nu-\mu}=0}$. The
%difference $\Delta_{mn} = F[p]-F[p_{-mn}]$ for all pairs composed out of
%slowest 25 modes of each protein, as well as their rank-ordered distribution is
%given in Fig.~\ref{fig:pairs}. 
We found that the highest-impact mode pairs are 1-3 for AncCR and AncGR1; and
1-7 and 2-6 for AncGR2.  Spatial fluctuations associated with these mode pairs
coincide with helices 7 and 10, along with the loop region preceeding helix-7.
Indeed, these helices form part of the ligand binding pocket, while the loop
before helix-7 is where the two X mutations are located. We additionally
observed a region on helix-9 with high sequence conservation score also involved
in mode coupling which, to our knowledge, has not been highlighted in earlier
studies.

%% Interestingly, locations on the protein structure where {\em %   both} modes
%display large displacement amplitude statistically % correlate with some of the
%aminoacids whose functional importance is % revealed by mutation studies (see
%Fig.~\ref{fig:dotproducts} % below). 

%% \begin{figure}[] % %\hspace*{-1cm} % \centering %
%\includegraphics[scale=0.25]{coupled_matr.eps} % \caption{Pairwise mode
%coupling scores for slowest 25 modes of the %   three members of GR protein
%lineage.} % \label{fig:pairs} % \end{figure}

\begin{figure}[b] \centering
\includegraphics[width=3in]{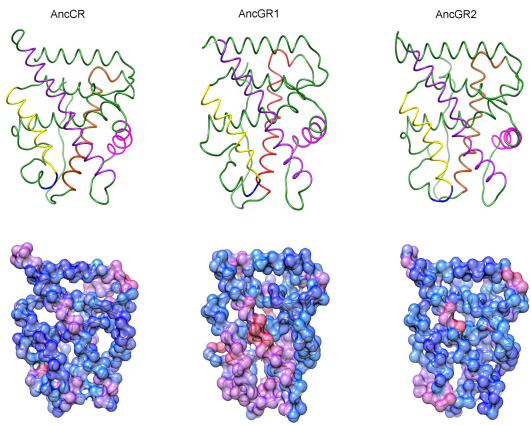} \caption{(a)
Cartoon representation of the three studies proteins. Helices 3 (29-54),7 (108-125), and 10 (180-210) are
shown in red, yellow, and purple, respectively. Loop region preceeding helix 7
is colored in blue with activation-function helix (AF-h)(220-232) in magenta. Helices
3,7, and 10 alongside with helix 1 (not shown) form a part of ligand binding
pocket. AF-h is essential for transcriptional activity. (b) Mode-coupling scores
mapped onto the corresponding protein structures where hotter colors represent
higher scores. Significant activity is observed on helices 7 and 10, and around
loop regions.}
\label{fig:js_mapped} \end{figure}

We next performed the analysis outlined in Section~\ref{sec:method} in order to
derive a mode-coupling score for each aminoacid. The resulting score vectors
obtained over the full data set (four trajectories) separately for each member
of the GR family are 
%plotted {\em vs} residue-ID in Fig.~\ref{fig:js}, and 
shown as a heat map superimposed onto the proteins' three-dimensional structure
in Fig.~\ref{fig:js_mapped}. We observed that the loop (100-110) preceeding
helix-7 yields considerably high scores in all proteins, despite the fact that
this loop and the nearby helix-7 exhibit the largest structural variability
between AncGR1 and AncGR2~\cite{ortlund2007crystal}. Furthermore, the same
region also accommodates 4 of the 6 (XYZ-)mutations mentioned above. These
observations hint at the relevance of mode coupling to the evolutionary history
of function in the GR protein's lineage, which we investigate below in further
detail.

%% \begin{figure}[] % %\hspace*{-1cm} % \centering %
%\includegraphics[scale=0.16]{js_ultimate_all_combined.eps} %
%\caption{Normalized JS distances for 2Q3Y, 3RY9, and 3GN8. 0.3 offset %   is
%applied to distinguish between different structures. Red dots %   show all
%mutations between successive evolutionary %   steps. XYZ-mutations are labeled
%as black dots. Green dots represent % Q114L/M197I mutations. Ligand binding
%pocket forming helices 1,3,7, %   and 10 (left to right) are also shown with
%faded %   blue. Activation-function helix is shown in turquoise. } %
%\label{fig:js} % \end{figure}

Note that, the location of the X-mutation S106P consistently has one of the
highest scores in all proteins. Considering that S106P {\em alone} decreases
activation in AncGR indepedent of the ligand type~\cite{ortlund2007crystal}
suggests that the mechanism underlying the activity loss is mechanical in
origin, rather than biochemical (to which the present method is insensitive).
The opposite is true for the second X mutation L111Q which recovers cortisol
specificity~\cite{ortlund2007crystal} by allowing formation of a hydrogen bond
with cortisol. Mode-coupling score of location 111 shows no significant
deviation from the mean.  On the other hand, the synthetic mutations Q114L/M197I
in AncGR1 - that also recover cortisol specificity {\em and} disrupt
communication between cortisol binding and transcriptional activity - coincide
with the two mode-coupling peaks in AncGR1 located on helix-7 and helix-10.

Complementing these observations, an objective evaluation of the correlation
between mode-coupling scores and the AncGR1 $\rightarrow$ AncGR2 mutation set is
desirable. For this purpose, we use the recall analysis where mutation sites
under consideration are labelled as the target set and their rankings are
inspected in the full residue list sorted according to mode-coupling scores. The
result is presented as a recall curve which is a plot of the fraction of the
target set elements ($y$-coordinate) observed in a given fraction
($x$-coordinate) of the list picked from the top.  In absence of correlation
between the target set and the scoring function, one expects to see the recall
to remain on the diagonal upto statistical fluctuations. A recall curve
remaining significantly above the diagonal indicates a positive correlation,
since it reflects the fact that the aminoacids in the target set come with
higher-than-average scores.

\begin{figure}[] 
%\hspace*{-1cm}
\centering
\includegraphics[scale=0.20]{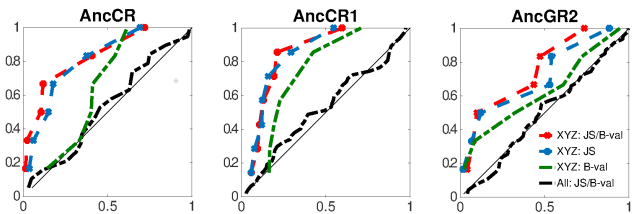} \caption{Recall analysis
for AncCR, AncGR1, AncGR2. For each protein, mode-coupling scores and B-values
(rms amplitude of C$_\alpha$ fluctuations) were used for ranking residues.
Target residues were set to be XYZ-mutation locations or the locations of all
mutations that occured between two evolutionary steps. All figures show a
significant positive bias towards XYZ-mutations when mode-coupling scores are
used. This pattern is lost when B-values are used for scoring.}
\label{fig:recall} \end{figure}

Fig.~\ref{fig:recall} shows the outcome of the recall analysis for the three
proteins in the GR family. In all cases, we observe no visible correlation
between mode-coupling scores and the complete set of mutations accompanying each
evolutionary step. Focusing on the function changing XYZ-mutations only, we
first note an overall positive correlation with B-values (variance of
fluctuations around equilibrium for each $C_{\alpha}$ atom), due to the fact
that these mutations are located mostly on the loop regions. It is striking that
the mode-coupling based ordering yields better recall values in all three
proteins. We furthermore observe that the recall performances improve slightly
when mode-coupling scores are divided by the B-values in order to factor out the
bias mentioned above. This is the central result of the present work which,
together with a similar observation on myosin II~\cite{varol2014mode}, lends
support to the thesis that coupling between vibrational modes is a key physical
mechanism in protein function.

\subsection{Robustness {\em wrt} data acquisition period} \label{sec:robustness}

Since most proteins carry out their function in time scales beyond the reach of
computer simulations, it is natural to ask how sensitive above results are to
the simulation time window. We investigated the robustness of our findings by
re-analyzing the data in varying time intervals. To this end, we divided each
trajectory into $N_T$ fragments of $T$ ns each ($T=1,2,4,8,\dots,128$) and
calculated mode-coupling scores by using each fragment separately. We then
compared score vectors for each interval pair with identical lengths by
measuring the angle between them. This was done by evaluating the dot product of
the two vectors after setting their mean to zero (by a \begin{figure}[]

\includegraphics[width=3in]{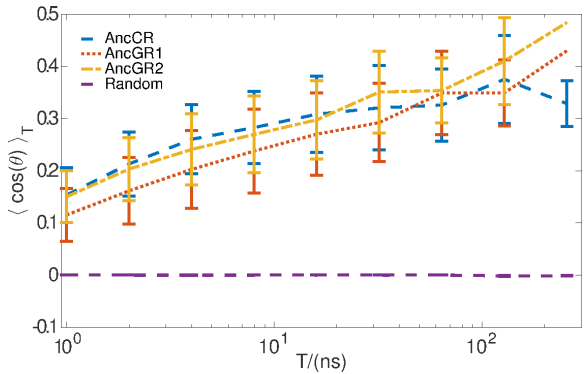} \caption{Self consistency
of the score vectors increases with the length of the MD trajectory.
$\langle\cos\theta\rangle_T$ is the average value of the dot product of two
normalized score vectors obtained from different time windows of size $T$.}
\label{fig:consistency} \end{figure} constant shift) and rescaling them to unit
length. The mean $\langle \cos{\theta}\rangle_T$ and the standard devitation
$\sigma_T$ of the obtained dot products were recorded separately for each
interval length $T$.  Results shown in Fig.~\ref{fig:consistency} confirm that
the analysis detailed in section~\ref{sec:method} yields progressively more
consistent results with increasing $T$.

\subsection{Conclusion} \label{sec:conclusion} While it is natural from a
physical point of view to postulate that nonlinear effects mitigate
energy transfer within a protein\cite{piazza2009long}, precisely how
the nonlinearity observed in protein dynamics can be fruitfully
exploited to yield biologically relevant predictions is unclear. Even
the relevance of nonlinear effects to protein function is far from
being universally acknowledged. While part of the literature (such as
on discrete
breathers~\cite{piazza2014nonlinear,luccioli2011discrete,juanico2007discrete,piazza2008discrete})
attests to its importance, there is substantial amount of past and
recent work which investigate mechanisms of protein function within a
linear (harmonic) framework or at the level of principal component
analysis~\cite{bahar2005coarse,haliloglu1997gaussian,levy1982molecular,ma2005usefulness,rodgers2013modulation,mcleish2015dynamic}. By
demonstrating that functionally critical mutations along the
evolutionary descent which relates three ancestral proteins of the GR
family are highlighted in an analysis of the nonlinear contribution to
dynamics, the present work emphasizes the significance of
nonlinearity, in particular that {\it beyond} marginal anharmonicity,
to protein function.

The selective power of the mode-coupling analysis for functionally
relevant sites (in GR protein family reported here and in myosin II
earlier~\cite{varol2014mode}) is suggestive. However, it is also
evident from the data that not all known functional locations come
with high mode-coupling scores. Given the complexity of the system and
the multitude of factors beyond protein dynamics that play role in
functionality, this is only expected.  Applying the analysis on
carefully constructed toy nonlinear models may help clarify the
mechanistic role played by the aminoacids that score high in the
present analysis. Such information could be useful for characterizing
the proteins on which the current approach may be expected to be
successful in future.

\subsection{Acknowledgements} 
We thank B. Erman, D. Yuret, O. Keskin and A.  Erdo\u gan for
beneficial discussions, C.  At\i lgan for a critical reading, and
S. B. Ozkan for a stimulating seminar which motivated the present
work. We acknowledge support of T\" UB\. ITAK through the grant
MFAG-113F092.

%\section{Figures}

\bibliographystyle{unsrtnat} \bibliography{references} 
\begin{thebibliography}{46}
\providecommand{\natexlab}[1]{#1}
\providecommand{\url}[1]{\texttt{#1}}
\expandafter\ifx\csname urlstyle\endcsname\relax
  \providecommand{\doi}[1]{doi: #1}\else
  \providecommand{\doi}{doi: \begingroup \urlstyle{rm}\Url}\fi

\bibitem[Kumar et~al.(2000)Kumar, Ma, Tsai, Sinha, and
  Nussinov]{kumar2000folding}
Sandeep Kumar, Buyong Ma, Chung-Jung Tsai, Neeti Sinha, and Ruth Nussinov.
\newblock Folding and binding cascades: dynamic landscapes and population
  shifts.
\newblock \emph{Protein Science}, 9\penalty0 (1):\penalty0 10--19, 2000.

\bibitem[Tsai et~al.(1999)Tsai, Ma, and Nussinov]{tsai1999folding}
Chung-Jung Tsai, Buyong Ma, and Ruth Nussinov.
\newblock Folding and binding cascades: shifts in energy landscapes.
\newblock \emph{Proceedings of the National Academy of Sciences}, 96\penalty0
  (18):\penalty0 9970--9972, 1999.

\bibitem[Okazaki and Takada(2008)]{okazaki2008dynamic}
Kei-ichi Okazaki and Shoji Takada.
\newblock Dynamic energy landscape view of coupled binding and protein
  conformational change: induced-fit versus population-shift mechanisms.
\newblock \emph{Proceedings of the National Academy of Sciences}, 105\penalty0
  (32):\penalty0 11182--11187, 2008.

\bibitem[Kar et~al.(2010)Kar, Keskin, Gursoy, and Nussinov]{kar2010allostery}
Gozde Kar, Ozlem Keskin, Attila Gursoy, and Ruth Nussinov.
\newblock Allostery and population shift in drug discovery.
\newblock \emph{Current opinion in pharmacology}, 10\penalty0 (6):\penalty0
  715--722, 2010.

\bibitem[Brooks and Karplus(1983)]{brooks1983harmonic}
Bernard Brooks and Martin Karplus.
\newblock Harmonic dynamics of proteins: normal modes and fluctuations in
  bovine pancreatic trypsin inhibitor.
\newblock \emph{Proceedings of the National Academy of Sciences}, 80\penalty0
  (21):\penalty0 6571--6575, 1983.

\bibitem[Haliloglu et~al.(1997)Haliloglu, Bahar, and
  Erman]{haliloglu1997gaussian}
Turkan Haliloglu, Ivet Bahar, and Burak Erman.
\newblock Gaussian dynamics of folded proteins.
\newblock \emph{Physical review letters}, 79\penalty0 (16):\penalty0 3090,
  1997.

\bibitem[Yogurtcu et~al.(2009)Yogurtcu, Gur, and
  Erman]{yogurtcu2009statistical}
Osman~N Yogurtcu, Mert Gur, and Burak Erman.
\newblock Statistical thermodynamics of residue fluctuations in native
  proteins.
\newblock \emph{The Journal of chemical physics}, 130\penalty0 (9):\penalty0
  095103, 2009.

\bibitem[Ma(2005)]{ma2005usefulness}
Jianpeng Ma.
\newblock Usefulness and limitations of normal mode analysis in modeling
  dynamics of biomolecular complexes.
\newblock \emph{Structure}, 13\penalty0 (3):\penalty0 373--380, 2005.

\bibitem[Piazza and Sanejouand(2009)]{piazza2009long}
Francesco Piazza and Yves-Henri Sanejouand.
\newblock Long-range energy transfer in proteins.
\newblock \emph{Physical biology}, 6\penalty0 (4):\penalty0 046014, 2009.

\bibitem[Rodgers et~al.(2013)Rodgers, Townsend, Burnell, Jones, Richards,
  McLeish, Pohl, Wilson, and Cann]{rodgers2013modulation}
Thomas~L Rodgers, Philip~D Townsend, David Burnell, Matthew~L Jones, Shane~A
  Richards, Tom~CB McLeish, Ehmke Pohl, Mark~R Wilson, and Martin~J Cann.
\newblock Modulation of global low-frequency motions underlies allosteric
  regulation: demonstration in crp/fnr family transcription factors.
\newblock \emph{PLoS Biol}, 11\penalty0 (9):\penalty0 e1001651, 2013.

\bibitem[Kabak{\c{c}}{\i}o{\u{g}}lu et~al.(2010)Kabak{\c{c}}{\i}o{\u{g}}lu,
  Yuret, Gur, and Erman]{kabakcciouglu2010anharmonicity}
A~Kabak{\c{c}}{\i}o{\u{g}}lu, D~Yuret, M~Gur, and B~Erman.
\newblock Anharmonicity, mode-coupling and entropy in a fluctuating native
  protein.
\newblock \emph{Physical biology}, 7\penalty0 (4):\penalty0 046005, 2010.

\bibitem[Varol et~al.(2014)Varol, Yuret, Erman, and
  Kabak{\c{c}}{\i}o{\u{g}}lu]{varol2014mode}
Onur Varol, Deniz Yuret, Burak Erman, and Alkan Kabak{\c{c}}{\i}o{\u{g}}lu.
\newblock Mode coupling points to functionally important residues in myosin ii.
\newblock \emph{Proteins: Structure, Function, and Bioinformatics}, 82\penalty0
  (9):\penalty0 1777--1786, 2014.

\bibitem[Flory and Yoon(1974)]{flory1974moments}
PJ~Flory and DY~Yoon.
\newblock Moments and distribution functions for polymer chains of finite
  length. i. theory.
\newblock \emph{The Journal of Chemical Physics}, 61\penalty0 (12):\penalty0
  5358--5365, 1974.

\bibitem[Hyv{\"a}rinen et~al.(2004)Hyv{\"a}rinen, Karhunen, and
  Oja]{hyvarinen2004independent}
Aapo Hyv{\"a}rinen, Juha Karhunen, and Erkki Oja.
\newblock \emph{Independent component analysis}, volume~46.
\newblock John Wiley \& Sons, 2004.

\bibitem[Karplus and McCammon(1983)]{kurplus1983dynamics}
M~Karplus and JA~McCammon.
\newblock Dynamics of proteins: elements and function.
\newblock \emph{Annual review of biochemistry}, 52\penalty0 (1):\penalty0
  263--300, 1983.

\bibitem[Levy et~al.(1982)Levy, Perahia, and Karplus]{levy1982molecular}
Ronald~M Levy, David Perahia, and Martin Karplus.
\newblock Molecular dynamics of an $\alpha$-helical polypeptide: temperature
  dependence and deviation from harmonic behavior.
\newblock \emph{Proceedings of the National Academy of Sciences}, 79\penalty0
  (4):\penalty0 1346--1350, 1982.

\bibitem[Levitt et~al.(1985)Levitt, Sander, and Stern]{levitt1985protein}
Michael Levitt, Christian Sander, and Peter~S Stern.
\newblock Protein normal-mode dynamics: trypsin inhibitor, crambin,
  ribonuclease and lysozyme.
\newblock \emph{Journal of molecular biology}, 181\penalty0 (3):\penalty0
  423--447, 1985.

\bibitem[Doruker et~al.(2000)Doruker, Atilgan, and Bahar]{doruker2000dynamics}
Pemra Doruker, Ali~Rana Atilgan, and Ivet Bahar.
\newblock Dynamics of proteins predicted by molecular dynamics simulations and
  analytical approaches: Application to $\alpha$-amylase inhibitor.
\newblock \emph{Proteins: Structure, Function, and Bioinformatics}, 40\penalty0
  (3):\penalty0 512--524, 2000.

\bibitem[Manning and Sch{\"u}tze(1999)]{manning1999foundations}
Christopher~D Manning and Hinrich Sch{\"u}tze.
\newblock \emph{Foundations of statistical natural language processing}.
\newblock MIT press, 1999.

\bibitem[Kino et~al.(2009)Kino, Manoli, Kelkar, Wang, Su, and
  Chrousos]{kino2009glucocorticoid}
Tomoshige Kino, Irini Manoli, Sujata Kelkar, Yonghong Wang, Yan~A Su, and
  George~P Chrousos.
\newblock Glucocorticoid receptor (gr) $\beta$ has intrinsic,
  gr$\alpha$-independent transcriptional activity.
\newblock \emph{Biochemical and biophysical research communications},
  381\penalty0 (4):\penalty0 671--675, 2009.

\bibitem[Nicolaides et~al.(2010)Nicolaides, Galata, Kino, Chrousos, and
  Charmandari]{nicolaides2010human}
Nicolas~C Nicolaides, Zoi Galata, Tomoshige Kino, George~P Chrousos, and
  Evangelia Charmandari.
\newblock The human glucocorticoid receptor: molecular basis of biologic
  function.
\newblock \emph{Steroids}, 75\penalty0 (1):\penalty0 1--12, 2010.

\bibitem[Chrousos(2004)]{chrousos2004glucocorticoid}
George~P Chrousos.
\newblock The glucocorticoid receptor gene, longevity, and the complex
  disorders of western societies.
\newblock \emph{The American journal of medicine}, 117\penalty0 (3):\penalty0
  204--207, 2004.

\bibitem[Clark~JK(1992)]{clark}
O'Malley~BW Clark~JK, Schrader~WT.
\newblock \emph{Mechanism of steroid hormones}, pages 35--90.
\newblock Philedelphia: WB Sanders Co., 1992.

\bibitem[Zhou and Cidlowski(2005)]{zhou2005human}
Junguo Zhou and John~A Cidlowski.
\newblock The human glucocorticoid receptor: one gene, multiple proteins and
  diverse responses.
\newblock \emph{Steroids}, 70\penalty0 (5):\penalty0 407--417, 2005.

\bibitem[Barnes(1998)]{barnes1998anti}
Peter~J Barnes.
\newblock Anti-inflammatory actions of glucocorticoids: molecular mechanisms.
\newblock \emph{Clinical science}, 94\penalty0 (6):\penalty0 557--572, 1998.

\bibitem[Sapolsky et~al.(2000)Sapolsky, Romero, and
  Munck]{sapolsky2000glucocorticoids}
Robert~M Sapolsky, L~Michael Romero, and Allan~U Munck.
\newblock How do glucocorticoids influence stress responses? integrating
  permissive, suppressive, stimulatory, and preparative actions 1.
\newblock \emph{Endocrine reviews}, 21\penalty0 (1):\penalty0 55--89, 2000.

\bibitem[Carson et~al.(2014)Carson, Luz, Suen, Montrose, Zink, Ruan, Cheng,
  Cole, Adrian, Kohlman, et~al.]{carson2014glucocorticoid}
Matthew~W Carson, John~G Luz, Chen Suen, Chahrzad Montrose, Richard Zink,
  Xiaoping Ruan, Christine Cheng, Harlan Cole, Mary~D Adrian, Dan~T Kohlman,
  et~al.
\newblock Glucocorticoid receptor modulators informed by crystallography lead
  to a new rationale for receptor selectivity, function, and implications for
  structure-based design.
\newblock \emph{Journal of medicinal chemistry}, 57\penalty0 (3):\penalty0
  849--860, 2014.

\bibitem[Li et~al.(2005)Li, Suino, Daugherty, and Xu]{li2005structural}
Yong Li, Kelly Suino, Jennifer Daugherty, and H~Eric Xu.
\newblock Structural and biochemical mechanisms for the specificity of hormone
  binding and coactivator assembly by mineralocorticoid receptor.
\newblock \emph{Molecular cell}, 19\penalty0 (3):\penalty0 367--380, 2005.

\bibitem[Bentley(1998)]{bentley1998comparative}
Peter~John Bentley.
\newblock \emph{Comparative vertebrate endocrinology}.
\newblock Cambridge University Press, 1998.

\bibitem[Yang and Young(2009)]{yang2009mineralocorticoid}
Jun Yang and Morag~J Young.
\newblock The mineralocorticoid receptor and its coregulators.
\newblock \emph{Journal of molecular endocrinology}, 43\penalty0 (2):\penalty0
  53--64, 2009.

\bibitem[Bridgham et~al.(2009)Bridgham, Ortlund, and
  Thornton]{bridgham2009epistatic}
Jamie~T Bridgham, Eric~A Ortlund, and Joseph~W Thornton.
\newblock An epistatic ratchet constrains the direction of glucocorticoid
  receptor evolution.
\newblock \emph{Nature}, 461\penalty0 (7263):\penalty0 515--519, 2009.

\bibitem[Ortlund et~al.(2007)Ortlund, Bridgham, Redinbo, and
  Thornton]{ortlund2007crystal}
Eric~A Ortlund, Jamie~T Bridgham, Matthew~R Redinbo, and Joseph~W Thornton.
\newblock Crystal structure of an ancient protein: evolution by conformational
  epistasis.
\newblock \emph{Science}, 317\penalty0 (5844):\penalty0 1544--1548, 2007.

\bibitem[Harms and Thornton(2014)]{harms2014historical}
Michael~J Harms and Joseph~W Thornton.
\newblock Historical contingency and its biophysical basis in glucocorticoid
  receptor evolution.
\newblock \emph{Nature}, 512\penalty0 (7513):\penalty0 203--207, 2014.

\bibitem[Glembo et~al.(2012)Glembo, Farrell, Gerek, Thorpe, and
  Ozkan]{glembo2012collective}
Tyler~J Glembo, Daniel~W Farrell, Z~Nevin Gerek, MF~Thorpe, and S~Banu Ozkan.
\newblock Collective dynamics differentiates functional divergence in protein
  evolution.
\newblock \emph{PLoS computational biology}, 8\penalty0 (3):\penalty0 e1002428,
  2012.

\bibitem[Case et~al.(2014)Case, Babin, Berryman, Betz, Cai, Cerutti,
  Cheatham~Iii, Darden, Duke, Gohlke, et~al.]{case2014amber}
DA~Case, V~Babin, Josh Berryman, RM~Betz, Q~Cai, DS~Cerutti, TE~Cheatham~Iii,
  TA~Darden, RE~Duke, H~Gohlke, et~al.
\newblock Amber 14.
\newblock 2014.

\bibitem[Cornell et~al.(1995)Cornell, Cieplak, Bayly, Gould, Merz, Ferguson,
  Spellmeyer, Fox, Caldwell, and Kollman]{cornell1995second}
Wendy~D Cornell, Piotr Cieplak, Christopher~I Bayly, Ian~R Gould, Kenneth~M
  Merz, David~M Ferguson, David~C Spellmeyer, Thomas Fox, James~W Caldwell, and
  Peter~A Kollman.
\newblock A second generation force field for the simulation of proteins,
  nucleic acids, and organic molecules.
\newblock \emph{Journal of the American Chemical Society}, 117\penalty0
  (19):\penalty0 5179--5197, 1995.

\bibitem[Wang et~al.(2004)Wang, Wolf, Caldwell, Kollman, and
  Case]{wang2004development}
Junmei Wang, Romain~M Wolf, James~W Caldwell, Peter~A Kollman, and David~A
  Case.
\newblock Development and testing of a general amber force field.
\newblock \emph{Journal of computational chemistry}, 25\penalty0 (9):\penalty0
  1157--1174, 2004.

\bibitem[Wang et~al.(2006)Wang, Wang, Kollman, and Case]{wang2006automatic}
Junmei Wang, Wei Wang, Peter~A Kollman, and David~A Case.
\newblock Automatic atom type and bond type perception in molecular mechanical
  calculations.
\newblock \emph{Journal of molecular graphics and modelling}, 25\penalty0
  (2):\penalty0 247--260, 2006.

\bibitem[Roh et~al.(2005)Roh, Novikov, Gregory, Curtis, Chowdhuri, and
  Sokolov]{roh2005onsets}
JH~Roh, VN~Novikov, RB~Gregory, JE~Curtis, Z~Chowdhuri, and AP~Sokolov.
\newblock Onsets of anharmonicity in protein dynamics.
\newblock \emph{Physical review letters}, 95\penalty0 (3):\penalty0 038101,
  2005.

\bibitem[Hayward et~al.(1995)Hayward, Kitao, and
  G{\=o}]{hayward1995harmonicity}
Steven Hayward, Akio Kitao, and Nobuhiro G{\=o}.
\newblock Harmonicity and anharmonicity in protein dynamics: a normal mode
  analysis and principal component analysis.
\newblock \emph{Proteins: Structure, Function, and Bioinformatics}, 23\penalty0
  (2):\penalty0 177--186, 1995.

\bibitem[Piazza(2014)]{piazza2014nonlinear}
Francesco Piazza.
\newblock Nonlinear excitations match correlated motions unveiled by nmr in
  proteins: a new perspective on allosteric cross-talk.
\newblock \emph{Physical biology}, 11\penalty0 (3):\penalty0 036003, 2014.

\bibitem[Luccioli et~al.(2011)Luccioli, Imparato, Lepri, Piazza, and
  Torcini]{luccioli2011discrete}
Stefano Luccioli, Alberto Imparato, Stefano Lepri, Francesco Piazza, and
  Alessandro Torcini.
\newblock Discrete breathers in a realistic coarse-grained model of proteins.
\newblock \emph{Physical biology}, 8\penalty0 (4):\penalty0 046008, 2011.

\bibitem[Juanico et~al.(2007)Juanico, Sanejouand, Piazza, and
  De~Los~Rios]{juanico2007discrete}
Brice Juanico, Y-H Sanejouand, Francesco Piazza, and Paolo De~Los~Rios.
\newblock Discrete breathers in nonlinear network models of proteins.
\newblock \emph{Physical review letters}, 99\penalty0 (23):\penalty0 238104,
  2007.

\bibitem[Piazza and Sanejouand(2008)]{piazza2008discrete}
Francesco Piazza and Yves-Henri Sanejouand.
\newblock Discrete breathers in protein structures.
\newblock \emph{Physical biology}, 5\penalty0 (2):\penalty0 026001, 2008.

\bibitem[Bahar and Rader(2005)]{bahar2005coarse}
Ivet Bahar and AJ~Rader.
\newblock Coarse-grained normal mode analysis in structural biology.
\newblock \emph{Current opinion in structural biology}, 15\penalty0
  (5):\penalty0 586--592, 2005.

\bibitem[McLeish et~al.(2015)McLeish, Cann, and Rodgers]{mcleish2015dynamic}
Tom~CB McLeish, Martin~J Cann, and Thomas~L Rodgers.
\newblock Dynamic transmission of protein allostery without structural change:
  Spatial pathways or global modes?
\newblock \emph{Biophysical journal}, 2015.

\end{thebibliography}
\end{document}